\documentclass[twocolumn,floatfix]{revtex4}

\usepackage{graphics}

\begin{document}

\title{Surface spontaneous parametric down-conversion}

\author{Jan Pe\v{r}ina Jr., Anton\'{\i}n Luk\v{s}, Ond\v{r}ej Haderka}
\affiliation{Joint Laboratory of Optics of Palack\'{y} University
and Institute of Physics of Academy of Sciences of the Czech
Republic, 17. listopadu 50A, 772 07 Olomouc, Czech Republic}
\author{Michael Scalora}
\affiliation{Charles M. Bowden Research Center, RD\&EC, Redstone
 Arsenal, Bldg 7804, Alabama 35898-5000, USA}

\begin{abstract}
Surface spontaneous parametric down-conversion is predicted as a
consequence of continuity requirements for electric- and
magnetic-field amplitudes at a discontinuity of $ \chi^{(2)} $
nonlinearity. A generalization of the usual two-photon spectral
amplitude is suggested to describe this effect. Examples of
nonlinear layered structures and periodically-poled nonlinear
crystals show that surface contributions to spontaneous
down-conversion can be important.
\end{abstract}


\maketitle

When studying the process of second-harmonic generation under
considerable phase mismatch more than thirty years ago, the
generation of second-harmonic field from a boundary between two
homogeneous media that differ by values of $ \chi^{(2)} $
nonlinearity has been discovered
\cite{Bloembergen1962,Bloembergen1969}. The surface
second-harmonic field arises here as a consequence of continuity
requirements for projections of electric- and magnetic-field
vector amplitudes into the plane of the boundary. Physically, a
pumping field at frequency $\omega$ creates a step profile of
nonlinear polarization at frequency $2 \omega$ and with wave
vector $ 2 k(\omega)$ that becomes the source of the usual volume
second-harmonic field. The wave vector of the surface
second-harmonic field is $ k(2\omega) $ in agreement with
dispersion properties of the nonlinear material. This effect is
even found in nonlinear media with negative index of refraction as
the numerical solution of nonlinear Maxwell equations revealed in
\cite{Roppo2007}. The studied parametric effect should be
distinguished from resonant surface second-harmonic generation.

Spontaneous parametric down-conversion (SPDC) \cite{Mandel1995}
belongs together with second-harmonic generation to $ \chi^{(2)} $
processes. This poses the question about surface effects in SPDC.
In volume SPDC, photon pairs are generated from the vacuum state,
due to quantum fluctuations (or quantum noise) inherent in this
state. In this case, a nonlinear material responds to the presence
of optical fields through quantum nonlinear polarization that acts
as a source of new fields. In a close vicinity of the boundary,
the interacting fields as well as the nonlinear polarization are
modified in order to comply with natural fields' continuity
requirements at the boundary. This results in the generation of
additional photon pairs from the area of the boundary (several
wavelengths thick) that constitute surface SPDC.

Our study of surface SPDC is organized as follows. Nonlinear
Heisenberg equations are derived first to treat SPDC inside the
nonlinear medium. Nonlinear corrections to electric- and
magnetic-field amplitudes occur naturally at boundaries and give
additional, i.e. surface, contributions to SPDC. Subsequently, the
derivation of quantities characterizing the emitted photon pairs
is addressed. Finally, two important examples are discussed.

Adopting the quantization of energy flux
\cite{Huttner1990,Luks2002} we describe the process of SPDC
involving the signal, idler, and pump fields by the Heisenberg
equations with an appropriate interaction momentum operator $
\hat{G}_{\rm int} $ \cite{Huttner1990}:
\begin{eqnarray}   
 \hat{G}_{\rm int}(z) &=& \frac{4 \epsilon_0 d_{\rm eff} {\cal A}}{ \sqrt{2\pi}}
  \sum_{\alpha,\beta,\gamma=F,B} \int d\omega_s \int d\omega_i
 \nonumber \\
 & &
  \hspace{-16mm}
  \left[ E^{(-)}_{p_\alpha}(z,\omega_s+\omega_i) \hat{E}^{(+)}_{s_\beta}(z,\omega_s)
  \hat{E}^{(+)}_{i_\gamma}(z,\omega_i)
  + {\rm h.c.} \right] .
\label{1}
\end{eqnarray}
The positive-frequency part of an electric-field amplitude  $
\hat{E}^{(+)}_{m_\alpha} $ can be expressed using annihilation
operator $ \hat{a}_{m_\alpha} $ as follows ($ m=p,s,i $; $ \alpha
= F,B $):
\begin{eqnarray} 
 \hat{E}_{m_\alpha}^{(+)}(z,\omega_m) = i \sqrt{ \frac{
  \hbar\omega_m }{ 2\epsilon_0 c{\cal A} n_m(\omega_m) } }
  \hat{a}_{m_\alpha}(z,\omega_m) ;
\label{2}
\end{eqnarray}
$ \hat{E}^{(-)}_{m_\alpha} = (\hat{E}^{(+)}_{m_\alpha})^\dagger $.
Subscript $ F $ ($ B $) indicates a field propagating forward
(backward), i.e. along $ +z $ ($ -z $) axis. Symbol $ \epsilon_0 $
means permittivity of vacuum, $ d_{\rm eff} $ is effective
nonlinear coefficient, $ {\cal A} $ transverse area of the fields,
$ c $ speed of light in vacuum, and $ {\rm h.c.} $ replaces the
hermitian-conjugated terms. Symbol $ k_{m_\alpha} $ is a wave
vector, $ \omega_m $ frequency, and $ n_m $ index of refraction of
field $ m_\alpha $.

The Heisenberg equations, e.g., for the signal-field operators $
\hat{a}_{s_\alpha}(z,\omega_s) $ can then be derived assuming
equal-space commutation relations \cite{PerinaJr2000}:
\begin{eqnarray}    
 \frac{ d\hat{a}_{s_\alpha}(z,\omega_s) }{dz} &=& i k_{s_\alpha}(\omega_s)
  \hat{a}_{s_\alpha}(z,\omega_s)  \nonumber \\
 & & \hspace{-2cm} + \sum_{\beta,\gamma=F,B}
  \int d\omega_i g(\omega_s,\omega_i) E^{(+)}_{p_\beta}(0,\omega_s+\omega_i)
  \nonumber \\
 & & \hspace{-2cm} \times \exp[ik_{p_\beta}(\omega_s+\omega_i)z]
  \hat{a}_{i_\gamma}^\dagger(z,\omega_i), \;\; \alpha=F,B;
\label{3}
\end{eqnarray}
Coupling constant $ g $, $ g(\omega_s,\omega_i) = 2i d_{\rm eff}
\sqrt{\omega_s \omega_i} / (c \sqrt{2\pi} $ $ \sqrt{n_s(\omega_s)
n_i(\omega_i)} ) $, is linearly proportional to nonlinear
coefficient $ d_{\rm eff} $.

The solution of Eq.~(\ref{3}) for annihilation operators $
\hat{a}_{s_\alpha}(z,\omega_s) $ up to the first power of $ g $
gives us the formula for operator $ \hat{E}^{(+)}_{s_\alpha} $
defined in Eq.~(\ref{2}):
\begin{eqnarray}   
 \hat{E}_{s_\alpha}^{(+)}(z,\omega_s) &=& i \sqrt{ \frac{
  \hbar\omega_s}{ 2\epsilon_0 c {\cal A} n_s(\omega_s) } }
  \exp[ ik_{s_\alpha}(\omega_s) z ]
  \nonumber \\
 & & \hspace{-2.3cm} \times \Bigl[ \hat{a}_{s_\alpha}(0,\omega_s) + \sum_{\beta,\gamma=F,B}
   \int d\omega_i g(\omega_s,\omega_i) \nonumber \\
 & & \hspace{-2.3cm} \times E^{(+)}_{p_\beta}(0,\omega_s+\omega_i) \exp[i\Delta
    k_{\beta,\alpha\gamma}(\omega_s,\omega_i) z/2]
   \nonumber \\
 & & \hspace{-2.3cm} \times z \,
   {\rm sinc} [\Delta k_{\beta,\alpha\gamma}(\omega_s,\omega_i) z/2]
   \hat{a}_{i_\gamma}^\dagger(0,\omega_i) \Bigr]; \;\; \alpha=F,B;
\label{4}
\end{eqnarray}
$ {\rm sinc}(x) = \sin(x)/x $ and $ \Delta
k_{\beta,\alpha\gamma}(\omega_s,\omega_i) =
k_{p_\beta}(\omega_s+\omega_i) - k_{s_\alpha}(\omega_s) -
k_{i_\gamma}(\omega_i) $.

The positive-frequency magnetic-field amplitude operators $
\hat{H}_{s_\alpha}^{(+)} $ can be derived using the formula $
H_{s_\alpha}^{(+)}(z,\omega_s) = - i/(\omega_s\mu_0) \partial
E_{s_\alpha}^{(+)}(z,\omega_s) / \partial z $ ($ \mu_0 $ denotes
permeability of vacuum) provided that the electric-field
[magnetic-field] amplitude $ E_{s_\alpha} $ [$ H_{s_\alpha} $] is
polarized along $ +x $ [$ +y $] axis. The obtained operator $
\hat{H}_{s_\alpha}^{(+)} $ can be decomposed into two parts
denoted as $ \hat{H}^{(+)\rm Fr} _{s_\alpha} $ and $
\hat{H}^{(+)\rm nFr} _{s_\alpha} $:
\begin{eqnarray}   
 \hat{H}_{s_\alpha}^{(+)}(z,\omega_s) &=&
  \hat{H}_{s_\alpha}^{(+){\rm Fr}}(z,\omega_s) +
  \hat{H}_{s_\alpha}^{(+){\rm nFr}}(z,\omega_s), \nonumber \\
 & &
\label{5} \\
 \hat{H}_{s_\alpha}^{(+){\rm Fr}}(z,\omega_s) &=&
 \frac{k_{s_\alpha}(\omega_s)}{\omega_s\mu_0}
 \hat{E}_{s_\alpha}^{(+)}(z,\omega_s),
\label{6} \\
 \hat{H}_{s_\alpha}^{(+){\rm nFr}}(z,\omega_s) &=& \sqrt{ \frac{ \hbar c
  n_s(\omega_s) }{ 2\mu_0 \omega_s {\cal A}}  }
  \sum_{\beta,\gamma=F,B} \int d\omega_i\, g(\omega_s,\omega_i) \nonumber \\
 & & \hspace{-2cm} \times
  E^{(+)}_{p_\beta}(\omega_s+\omega_i) \exp[ik_{p_\beta}(\omega_s+\omega_i) z] \nonumber \\
  & & \hspace{-2cm} \times  \exp[-ik_{i_\gamma}(\omega_i)z]
  \hat{a}^\dagger_{i_\gamma}(0,\omega_i) ,
   \hspace{5mm} \alpha=F,B .
\label{7}
\end{eqnarray}
By definition, the magnetic-field amplitude operator $
\hat{H}^{(+){\rm Fr}}_{s_\alpha}(z,\omega_s) $ is linearly
proportional to the electric-field amplitude operator $
\hat{E}^{(+)}_{s_\alpha}(z,\omega_s) $. The remaining
magnetic-field operator $ \hat{H}^{(+){\rm nFr}}_{s_\alpha} $ is
of purely nonlinear origin and the usual derivation of Fresnel
relations does not take it into account. Standard approaches to
nonlinear interactions thus do not involve this nonlinear term and
so they neglect surface effects. We note that the 'nonlinear'
magnetic-field operator $ \hat{H}^{(+){\rm nFr}}_{s_\alpha} $
occurs as a classical field amplitude also in the description of
stimulated parametric processes (e.g., in difference-frequency
generation) and yields surface contributions to these processes.

\begin{figure}  
\resizebox{0.6\hsize}{!}{\includegraphics{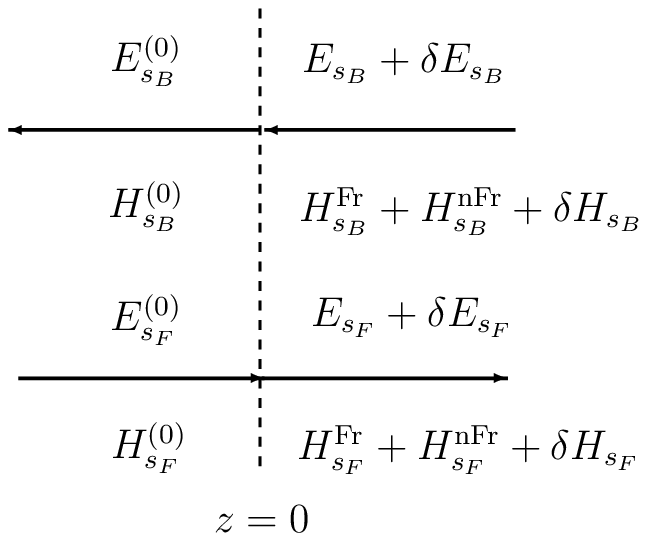}}
 \caption{Scheme showing electric and magnetic fields at the input boundary.
 For details, see the text.}
\label{fig1}
\end{figure}

The electric- and magnetic-field amplitudes $
E_{m_\alpha}(z,\omega_m) $ and $ H_{m_\alpha}(z,\omega_m) $
originating in the nonlinear interaction and written in
Eqs.~(\ref{4}) and (\ref{5}---\ref{7}) have to obey continuity
requirements at the input and output boundaries of the nonlinear
medium. We illustrate our approach to this problem considering the
signal field at the input boundary ($ z=0 $). Four electric and
magnetic fields are involved in the continuity requirements at
this boundary (see Fig.~\ref{fig1}): two at the linear left-hand
side [denoted by superscript $ (0) $] and two at the nonlinear
right-hand side. Because the magnetic-field amplitudes $
H_{s_F}^{} $ and $ H_{s_B}^{} $ inside the nonlinear medium have
also nonlinear contributions $ H_{s_F}^{\rm nFr} $ and $
H_{s_B}^{\rm nFr} $ given in Eq.~(\ref{7}) additional (surface)
amplitude corrections $ \delta E_{s_F} $ and $ \delta
E_{s_B}^{(0)} $ [together with $ \delta H_{s_F} $ and $ \delta
H_{s_B}^{(0)} $] in the fields leaving the boundary naturally
occur. The amplitude corrections $ \delta E_{s_B}^{(0)} $ and $
\delta H_{s_B}^{(0)} $ of the outgoing field outside the nonlinear
medium can be involved in the fields obeying Fresnel relations
\cite{Wolf1980} at the expense of introduction of fictitious
amplitude corrections $ \delta E_{s_B} $ and $ \delta H_{s_B} $ of
the field impinging at the boundary from its nonlinear side. A
detailed analysis then results in two equations for the surface
amplitude corrections of fields inside the nonlinear medium:
\begin{eqnarray}   
 0 &=& \delta E_{s_F}(0) -  \delta E_{s_B}(0),
  \nonumber \\
 0 &=& H_{s_F}^{\rm nFr}(0) + \delta H_{s_F}(0) +
  H_{s_B}^{\rm nFr}(0) - \delta H_{s_B}(0) .
\label{8}
\end{eqnarray}

The positive-frequency parts of surface amplitude-correction
operators $ \delta \hat{E}_{s_\alpha}^{(+)} $ and $ \delta
\hat{H}_{s_\alpha}^{(+)} $ occurring in the quantum form of
Eqs.~(\ref{8}) can be expressed using annihilation-operator
corrections $ \delta\hat{a}_{s_\alpha} $ similarly to the
corresponding amplitude operators $ \hat{E}_{s_\alpha}^{(+)} $ and
$ \hat{H}_{s_\alpha}^{(+)} $ in Eqs.~(\ref{2}) and (\ref{6}). The
solution of Eqs.~(\ref{8}) for $ \delta\hat{a}_{s_F} $ and $
\delta\hat{a}_{s_B} $ then takes the form:
\begin{eqnarray}  
 \delta a_{s_F}(0,\omega_s) &=& \delta a_{s_B}(0,\omega_s)=
 \frac{i}{k_s(\omega_s)}
  \sum_{\beta,\gamma=F,B}
  \nonumber \\
 & & \hspace{-15mm} \int d\omega_i\, g(\omega_s,\omega_i) E^{(+)}_{p_\beta}(0,\omega_s+\omega_i)
  \hat{a}^\dagger_{i_\gamma}(0,\omega_i).
\label{9}
\end{eqnarray}

Similar considerations appropriate for the output boundary leaves
us finally with an expression for operators $
\hat{a}_{s_\alpha}(L,\omega_s) $ valid up to the first power of $
g $ ($ L $ stands for the length of nonlinear medium):
\begin{eqnarray}    
 \hat{a}_{s_\alpha}(L,\omega_s) &=&
   \hat{a}_{s_\alpha}^{\rm free}(L,\omega_s) + \sum_{\beta,\gamma=F,B}
   \nonumber \\
  & & \hspace{-3cm} \int d\omega_i \,
   {\cal F}^s_{\alpha,\beta\gamma}(\omega_s,\omega_i)
   \hat{a}_{i_\gamma}^{{\rm free}\dagger}(L,\omega_i), \;\; \alpha=F,B.
\label{10}
\end{eqnarray}
Operators $ \hat{a}_{s_\alpha}^{{\rm free}}(L,\omega_s) $
correspond to free-field linear propagation, i.e. without
photon-pair generation. The idler-field amplitudes can be analyzed
along the same vein.

The generalized two-photon spectral amplitudes $ {\cal F}^s $ and
$ {\cal F}^i $ defined in Eq.~(\ref{10}) describe properties of a
generated photon pair and are composed of two contributions:
\begin{equation}   
 {\cal F}^{m}_{\alpha,\beta\gamma} = {\cal F}^{\rm vol}_{\alpha,\beta\gamma} + {\cal F}^{m,\rm
 surf}_{\alpha,\beta\gamma}; \;\; \alpha,\beta,\gamma=F,B.
\label{11}
\end{equation}
Two-photon spectral amplitude $ {\cal F}^{\rm vol} $ of the volume
contribution has the well-known form:
\begin{eqnarray} 
 {\cal F}^{\rm vol}_{\alpha,\beta\gamma}(\omega_s,\omega_i) &=& g(\omega_s,\omega_i)
  E^{(+)}_{p_\alpha}(0,\omega_s+\omega_i) \nonumber \\
 & & \hspace{-3cm} \times  \exp[ik_{p_\alpha}(\omega_s+\omega_i)L] \exp[-i\Delta k_{\alpha,\beta\gamma}
 (\omega_s,\omega_i)
   L/2] \nonumber \\
 & & \hspace{-3cm} \times  L \,
   {\rm sinc} [\Delta k_{\alpha,\beta\gamma}(\omega_s,\omega_i) L/2]; \;\; \alpha,\beta,\gamma=F,B .
\label{12}
\end{eqnarray}
On the other hand, surface contributions $ {\cal F}^{m,\rm surf} $
to the two-photon spectral amplitudes can be expressed as:
\begin{equation}   
 {\cal F}^{m,\rm surf}_{\alpha,\beta\gamma}(\omega_s,\omega_i) =
  {\cal V}_{\alpha,\beta\gamma}^m(\omega_s,\omega_i)
  {\cal F}^{\rm vol}_{\alpha,\beta\gamma}(\omega_s,\omega_i),
\label{13}
\end{equation}
where
\begin{equation}   
 {\cal V}_{\alpha,\beta\gamma}^m(\omega_s,\omega_i) = \frac{\Delta
  k_{\alpha,\beta\gamma}(\omega_s,\omega_i) }{ k_m(\omega_m) };
  \;\; \alpha,\beta,\gamma=F,B .
\label{14}
\end{equation}
The structure of surface contributions as described by the
two-photon amplitudes $ {\cal F}^{s,\rm surf} $ and $ {\cal
F}^{i,\rm surf} $ resembles that of the volume contribution as the
formula in Eq.~(\ref{13}) indicates. The physical interpretation
is as follows. At a boundary, the only restriction for photon-pair
generation is imposed by the conservation of energy. However, the
mutual interference of two-photon amplitudes originating at the
input and output boundaries leads to the result that resembles the
usual phase-matching conditions. We note that $ \lim_{L\rightarrow
0} {\cal F}^{m,\rm surf} = 0 $.

We further consider photon pairs with both photons propagating
forward and use operators $ \hat{a}_{m}(\omega_m) $ ($ m=s,i $)
defined outside the nonlinear medium. The joint signal-idler
photon-number density $ n(\omega_s,\omega_i) $ at the output plane
of the nonlinear medium is given as:
\begin{eqnarray}   
 n(\omega_s,\omega_i) &=& \left\langle
  \left[  \hat{a}_s^\dagger(\omega_s)
  \hat{a}_s(\omega_s) \hat{a}_i^\dagger (\omega_i)
  \hat{a}_i(\omega_i)  + {\rm h.c.}  \right]
  \right\rangle /2 . \nonumber \\
 & &
\label{15}
\end{eqnarray}
Symbol $ \langle \, \rangle $ denotes averaging over the initial
signal- and idler-field vacuum state. Introducing two-photon
spectral amplitudes $ \tilde{\cal F}^{s} $ and $ \tilde{\cal
F}^{i} $ (transmission coefficients $ t_m $ describe the output
boundary),
\begin{equation} 
 \tilde{\cal F}^{m}(\omega_s,\omega_i) = t_s(\omega_s) t_i(\omega_i)
  {\cal F}^{m}_{F,FF}(\omega_s,\omega_i), \,\, m=s,i ,
\label{16}
\end{equation}
we arrive at the following formula:
\begin{equation}  
 n(\omega_s,\omega_i) = {\rm Re} \{
  \tilde{\cal F}^{s*}(\omega_s,\omega_i)
  \tilde{\cal F}^{i}(\omega_s,\omega_i) \} .
\label{17}
\end{equation}
As this example illustrates, a generalization of the usual
formalism based on a two-photon spectral amplitude can be given
providing formulas for all physical quantities characterizing
photon pairs.

The volume interaction among the forward-propagating pump, signal,
and idler fields dominates in bulk nonlinear crystals several mm
long. According to our model, the surface contributions can be
approximately included into the usual formalism working with a
two-photon spectral amplitude $ \Phi^{\rm vol} $ (see, e.g.,
\cite{Keller1997,PerinaJr1999}) using the formal substitution:
\begin{eqnarray}   
 \Phi(\omega_s,\omega_i) & \longleftarrow & \sqrt{ 1+{\cal V}^{s}_{F,FF}
  (\omega_s,\omega_i) } \nonumber \\
 & &  \hspace{-5mm} \times \sqrt{ 1+{\cal V}^{i}_{F,FF}(\omega_s,\omega_i)
 } \, \Phi^{\rm vol}(\omega_s,\omega_i) ;
\label{18}
\end{eqnarray}
$ {\cal V} $ is defined in Eq.~(\ref{14}). If the nonlinear
interaction is perfectly phase-matched [$ \Delta
k_{F,FF}(\omega_s^0,\omega_i^0) = 0 $] the surface contributions
at central frequencies are zero.

Contrary to the bulk nonlinear crystals, surface SPDC cannot be
neglected in nonlinear layered structures composed of layers
typically several hundreds of nm long. In this case all possible
nonlinear interactions as described by the momentum operator $
\hat{G}_{\rm int} $ in Eq.~(\ref{1}) give appreciable
contributions. Fulfilment of phase-matching conditions is not
important here, because $ \Delta k l \ll \pi $ ($ l $ denotes a
typical length of one layer). Surface SPDC similarly as volume
SPDC from an individual layer is weak but both of them are highly
enhanced by constructive interference of fields from different
layers. A generalization of the presented theory to layered
structures is straightforward following the work presented in
\cite{PerinaJr2006,Centini2005}.

\begin{figure}  
\resizebox{0.95\hsize}{0.55\hsize}{\includegraphics{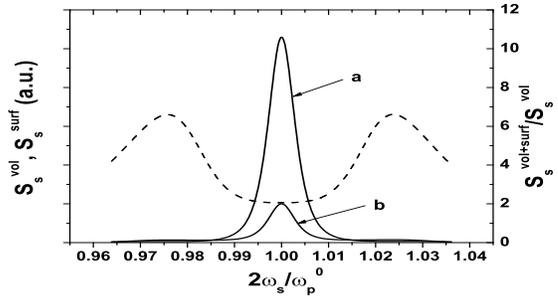}}
 \caption{Signal-field spectra $ S_s^{\rm vol} $ (solid curve denoted as a)
 and $ S_s^{\rm surf} $ (solid curve denoted as b) of volume and surface
 SPDC, respectively, and ratio $ S_s^{\rm vol+surf}/S_s^{\rm vol} $ of
 the spectra with ($ S_s^{\rm vol+surf} $) and
 without ($ S_s^{\rm vol} $) the inclusion of surface
 SPDC (dashed curve);
 $  S_s(\omega_s) = \hbar\omega_s \int d\omega_i
  \, n(\omega_s,\omega_i) $.}
\label{fig2}
\end{figure}

As an example, we consider a structure composed of 25 layers of
nonlinear GaN 117~nm thick that sandwich 24 linear layers of AlN
180~nm thick and studied previously in \cite{PerinaJr2006}. Volume
SPDC gives efficient photon-pair generation at degenerate signal-
and idler-field frequencies for the signal-field emission angle
14~deg \cite{PerinaJr2006} (see Fig.~\ref{fig2}) assuming a
normally incident pump field at $ \lambda_p = 664.5 $~nm and
s-polarized fields. Additional photon pairs originate in surface
SPDC. Their intensity is cca 20~\% of that coming from the volume.
However, both contributions are in phase and add constructively so
that the inclusion of surface SPDC roughly doubles the number of
emitted photon pairs in the spectral area of efficient photon-pair
generation (see Fig.~\ref{fig2}). Different contributions to
surface SPDC can be quantified using coefficients $ {\cal
V}^{m}_{\alpha,\beta\gamma} $ defined in Eq.~(\ref{14}). Whenever
the lengths of nonlinear layers are less or comparable to the
coherence length of the nonlinear process, we observe appreciable
contributions of surface terms. For example, the coherence length
equals approximately 1 $ \mu $m for GaN in our case.

\begin{figure}  
\resizebox{0.95\hsize}{0.55\hsize}{\includegraphics{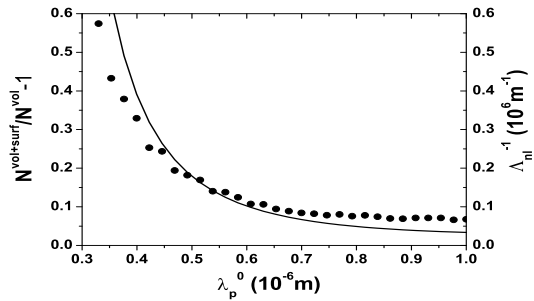}}
 \caption{Relative contribution $ N^{\rm vol+surf}/N^{\rm vol} -1 $
 of surface SPDC to
 the photon-pair generation rate (points $ \bullet $) and
 inverse $ \Lambda_{\rm nl}^{-1} $ of the optimum poling period giving
 quasi-phase-matching for $ \lambda_s^0=\lambda_i^0= 2 \lambda_p^0 $ (solid
 curve) as they depend on cw pump-field wavelength $ \lambda_p^0 $ in
 LiNbO$ {}_3 $ 5~mm long; $ N = \int d\omega_s \int d\omega_i
  \, n(\omega_s,\omega_i) $.}
\label{fig3}
\end{figure}

Surface effects give also an important contribution to photon-pair
generation rates in periodically-poled nonlinear materials with
sufficiently short poling periods. Here, as an example, we
consider frequency-degenerate SPDC in periodically-poled
LiNbO${}_3 $ with the optical axis perpendicular to the direction
of collinearly-propagating fields; their polarizations are
parallel to the optical axis. Whereas the surface effects
contribute to photon-pair generation rate $ N^{\rm vol+surf} $
only by several percent for the pump wavelength $ \lambda_p^0 = 1
$~$ \mu $m, the increase of photon-pair generation rate $ N $ by
50~\% is observed for $ \lambda_p^0 = 0.35 $~$ \mu $m (see
Fig.~\ref{fig3}). As the curves in Fig.~\ref{fig3} indicate the
relative contribution $ N^{\rm vol+surf} /N^{\rm vol} -1 $ of
surface terms is roughly proportional to the inverse $
1/\Lambda_{\rm nl} $ of poling period that is linearly
proportional to the density of surfaces per a unit length. We note
that domains shorter than 1~$ \mu $m can be fabricated using
light-induced domain engineering \cite{Sones2008}.

Surface SPDC occurs also in nonlinear wave-guiding structures,
i.e. under the condition of total reflection and presence of
evanescent waves. A detailed analysis has shown that the formulas
presented above remain valid also in this case provided that we
consider propagation constants $ \beta $ instead of wave vectors.
Coupling constant $ g $ then involves the overlap integral over
mode functions of the interacting fields. This is particularly
interesting for nonlinear photonic-band-gap fibers.

Surface SPDC is by no means restricted to 1D nonlinear structures:
even greater relative contributions are expected in 2D and 3D
nonlinear samples. Surface effects will also affect stimulated $
\chi^{(2)} $ processes like second-harmonic or second-subharmonic
generation when studied under comparable conditions.
Qualitatively, they will effectively enhance the nonlinearity.
This may be particularly interesting for squeezed-light
generation.

In conclusion, surface SPDC has been predicted. Generalized
signal- and idler-field two-photon spectral amplitudes have been
suggested to determine properties of emitted photon pairs. Surface
SPDC is important whenever strongly phase-mismatched nonlinear
interactions give considerable contributions. This occurs, e.g.,
in nonlinear layered structures or periodically-poled materials
where surface and volume contributions can be comparable. Surface
SPDC may affect optimum design of these structures that are
considered as promising versatile sources of photon pairs for
optoelectronics.

Support by the projects IAA100100713 of GA AV \v{C}R, COST 09026,
1M06002, and MSM6198959213 of the Czech Ministry of Education is
acknowledged.

\end{document}